\documentclass[aps,pra,reprint,superscriptaddress,showpacs]{revtex4-1}

\usepackage{graphicx}
\usepackage{amssymb,amsmath}
\usepackage{bm}
\usepackage[pdftex,breaklinks,colorlinks,citecolor=blue,linkcolor=blue,urlcolor=blue]{hyperref}
\usepackage{mathrsfs}

\begin{document}

\title{Three-body recombination near a narrow Feshbach resonance in $^6$Li}

\author{Jiaming Li}
\affiliation{School of Physics and Astronomy and Tianqin Research Center for Gravitational Physics, Sun Yat-Sen University,
Zhuhai, Guangdong, China 519082}
\affiliation{Department of Physics, Indiana University Purdue University Indianapolis, Indianapolis, IN 46202}
\author{Ji Liu}
\affiliation{Department of Physics, Indiana University Purdue University Indianapolis, Indianapolis, IN 46202}
\author{Le Luo}
\email[]{leluo@iupui.edu}
\affiliation{School of Physics and Astronomy and Tianqin Research Center for Gravitational Physics, Sun Yat-Sen University,
Zhuhai, Guangdong, China 519082}
\affiliation{Department of Physics, Indiana University Purdue University Indianapolis, Indianapolis, IN 46202}
\author{Bo Gao}
\email[]{bo.gao@utoledo.edu}
\affiliation{Department of Physics and Astronomy, Mailstop 111, University of Toledo, Toledo, OH 43606}

\date{\today}

\begin{abstract}

We experimentally measure, and theoretically analyze the three-atom recombination rate, $L_3$, around a narrow $s$ wave magnetic Feshbach resonance of $^6$Li-$^6$Li at 543.3 Gauss. By examining both the magnetic field dependence and especially the temperature dependence of $L_3$ over a wide range of temperatures from a few $\mu$K to above 200 $\mu$K, we show that three-atom recombination through a narrow resonance follows a universal behavior determined by the long-range van der Waals potential, and can be described by a set of rate equations in which three-body recombination proceeds via successive pairwise interactions. We expect the underlying physical picture to be applicable not only to narrow $s$ wave resonances, but also to resonances in nonzero partial waves, and not only at ultracold temperatures, but also at much higher temperatures.

\end{abstract}


\maketitle

Molecule formation through three-body recombination is one of the most fundamental chemical reactions as it pertains to the very origin of molecules \cite{Turk11,Forrey13} and their relative concentration to atomic species. It is also the key to understanding the initial stages of condensation where atoms form molecules, which further recombine with other atoms or molecules to grow into bigger molecules, clusters, and eventually to mesoscopic and macroscopic objects. As a reflection of the fundamental difficulties in quantum few-body systems, progress on three-body recombination has been excruciatingly slow. Fundamental questions such as the relative importance of direct (background or nonresonant) and indirect (successive pairwise or resonant) processes \cite{Greene14,Forrey15} seem as fresh as they were decades ago \cite{Wei97,Pack98}. Unlike deeply bound few-body bound states, for which large basis expansion works to a degree (see, e.g., \cite{Suzuki98}), three-body recombination occurs at much higher energies around the three-body breakup threshold where the number of open channels for most atoms other than helium goes to practically infinite, making standard numerical methods \cite{Suno02} impractical.

Cold-atom experiments have provided the experimental background for breakthroughs in few-body physics. In such experiments, two-body interaction can be precisely controlled via a Feshbach resonance (FR) \cite{Chin10}, and remarkably, manifestations of three-body recombination have become one of the most routinely measured quantities through trap loss. Vast amount of data thus generated has enabled considerable progress in few-body physics, first in elucidating the Efimov universality \cite{Nielsen01,BH06,Kraemer06,Greene10}, and more recently in discovery and exploration of the van der Waals universality (see, e.g., Refs.~\cite{Berninger11,Wang12,Wang12b,Naidon14,Blume15,Mestrom17,Johansen17}). Still, much of the progress has so far been limited to zero temperature, to broad $s$ wave FR's, and to the Efimov regime where the $s$ wave scattering lengths among the interacting particles are much greater than the ranges of interactions as measured by their corresponding van der Waals length scales. While experiments in other regimes are possible (see, e.g., Ref.~\cite{Hazlett12,Wang13}), they have not received as much attention partly due to the scarcity of the corresponding theories and partly due to the belief that such behaviors are not universal.

In this Letter, we first reassert that universal behaviors for few-atom and many-atom systems exist much beyond the zero temperature and beyond the $s$ wave, as first suggested some years ago \cite{Gao03}. They exist to 1 kelvin regime similar to the corresponding quantum-defect theory (QDT) for two-body interactions \cite{Gao08a,Gao05a}, and can be further extended to greater temperature regimes through multiscale QDT \cite{Gao16a}. Such broader-sense van der Waals universal behaviors can be mathematically rigorously defined in a way similar to the definitions of universal equations of states at the van der Waals length scale \cite{Gao04a,Gao05b,KG06}. They will be investigated as a part of a QDT for few-atom and many-atom systems. By expanding the region of universal behavior beyond the zero temperature and beyond a broad $s$ wave resonance, one will finally make the connection between studies of idealized few-body systems and real chemistry \cite{Forrey13,Greene14,Forrey15,Wei97,Pack98,Suno02}.
We take a step in this direction here by experimentally measure and theoretically analyze the three-atom recombination around a narrow $s$ wave magnetic FR of $^6$Li-$^6$Li around 543.3 Gauss. We show that at ultracold, but finite, temperatures, three-body recombination is dominated by the indirect process if there exists a narrow resonance within $k_B T$ above the threshold. We further show that the rate constant describing this successive pairwise process follows a universal behavior determined by the long-range van der Waals potential. An analytic formula is presented for the rate constant describing both its dependence on the temperature and its dependence on the resonance position, which in our case is tunable via a magnetic field.

\textit{Experiment:} We prepare a gas of $^6$Li atoms in the two lowest hyperfine states of $F=1/2,m_F=\pm1/2$ [labeled as $a$ ($+1/2$) and $b$ ($-1/2$) state, respectively] in a magneto-optical trap. The pre-cooled atoms are then transferred into a crossed-beam optical dipole trap (ODT) made by a fiber laser with 100 watt output. The bias magnetic field is quickly swept to 330 G to implement evaporative cooling~\cite{JL16,JL17}. A noisy radio-frequency pulse is then applied to prepare a 50:50 spin mixture. At 330 G, the trap potential can be lowered down to 0.1$\%$ of the full trap depth (the full trap depth is around 5.6 mK) to obtain a degenerate Fermi gas.  After that, the magnetic field is swept well above the narrow FR at 550 G to calibrate the temperature and the initial atom number $N_0$. Here the $s$-wave scattering length of $a$-$b$ state is close to the background scattering length of approximately  $0.96\beta_6$, for which the gas is weakly interacting (here $\beta_6 :=(2\mu C_6/\hbar^2)^{1/4}$ is the van der Waals length scale for $^6$Li-$^6$Li interaction \cite{Gao08a}).  The temperature of a weakly interacting Fermi gas is then measured by fitting the 1-D density profile with a finite temperature Thomas-Fermi distribution~\cite{luothesis}. To study the temperature dependence of three-body recombination rate, atom clouds are prepared in a temperature range between 4 $\mu$K and up to 225 $\mu$K by controlling the final trap depth and the evaporative cooling time.

To study three-body recombination rate around the narrow FR at 543.3 G, the magnetic field is fast swept from 550 G to a target field $B_t$ near the narrow resonance, where we hold atom cloud for a time duration $t$. To precisely locate the magnetic field, we record the magnets current during the holding time to monitor the fluctuation of the magnetic field.  After the holding period, the number of atoms left in the trap, $N(t)$, and the Gaussian width of the cloud, $\sigma_x(y,z)$, are extracted from the 2-D column density of the absorption images. To avoid the high column density induced error of the atom number, we turn off the optical trap after the holding period and take the absorption images of time-of-flight clouds.

\begin{figure}
\includegraphics[width=\columnwidth]{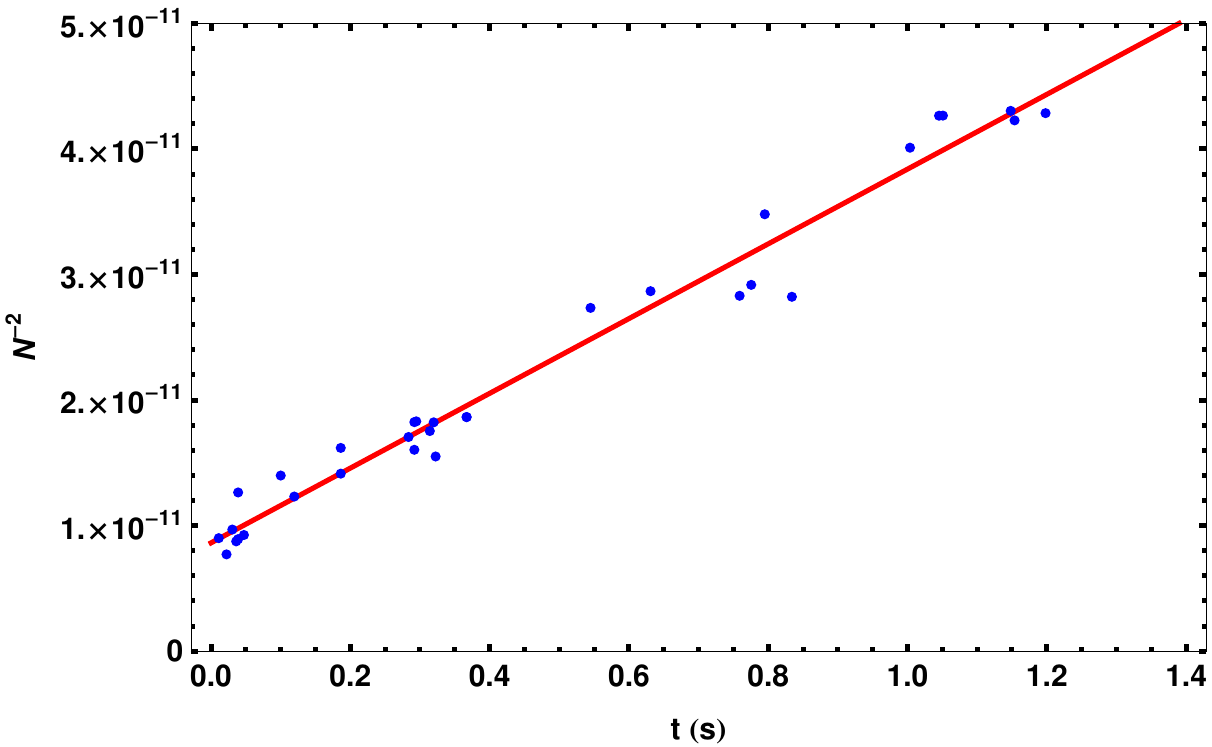}
\caption{The time-dependent $1/N^2$ where $N$ is the atom number left in the optical trap. The data is taken for a 70 $\mu$K cloud at the magnetic field 543.9 G. The fitting gives $L_3=2.455 \times 10^{-25}$ cm$^6$/s. }
\label{fig:Ntime}
\end{figure}

Our atomic vapor is a two-component thermal gas with $N_a$ atoms in state $a$ and $N_b$ atoms in state $b$. If the densities for atoms in states $a$ and $b$ start out the same, they will remain the same, namely $n_{a}=n_{b}=: n$, and decay with the same rate, by
\begin{equation}
\frac{dn}{dt} = -L_3 n^3,
\label{eq:L3def}
\end{equation}
where $L_3$ is the three-body recombination rate. The total atom number $N_a=N_b=:N$ is determined by integrating the density of the whole cloud, where we assume the profile is a Gaussian of the form $n(x,y,z)= n_0 \exp [-x^2/(\sigma_x^2)-y^2/(\sigma_y^2)-z^2/(\sigma_z^2)]$ with $n_0$ being the atom density at the center of the cloud. The integration gives us
\begin{equation}
\label{eq:3bodydecay}
\frac{dN(t)}{dt}=-\frac{L_3}{(2\sqrt3\pi)^3 \sigma_x^2\sigma_y^2\sigma_y^2}N^3(t) \;,
\end{equation}
implying that $1/N^2$ has a linear dependence on the holding time $t$ with
\begin{equation}
\label{eq:3bodyfit}
\frac{1}{N^2(t)}=\frac{2L_3}{(2\sqrt3\pi)^3 \sigma_x^2\sigma_y^2\sigma_y^2}t+\frac{1}{N^2(0)} \;.
\end{equation}
A typical time-dependent atom number data is shown in Fig.~\ref{fig:Ntime}, where the atom loss of a 70 $\mu$K cloud is taken at a holding field of 543.9 G. By fitting $1/N^2$ using Eq.~(\ref{eq:3bodyfit}), we extract $L_3$.

\begin{figure}
\begin{center}$
\begin{array}{c}
\includegraphics[width=2.5in]{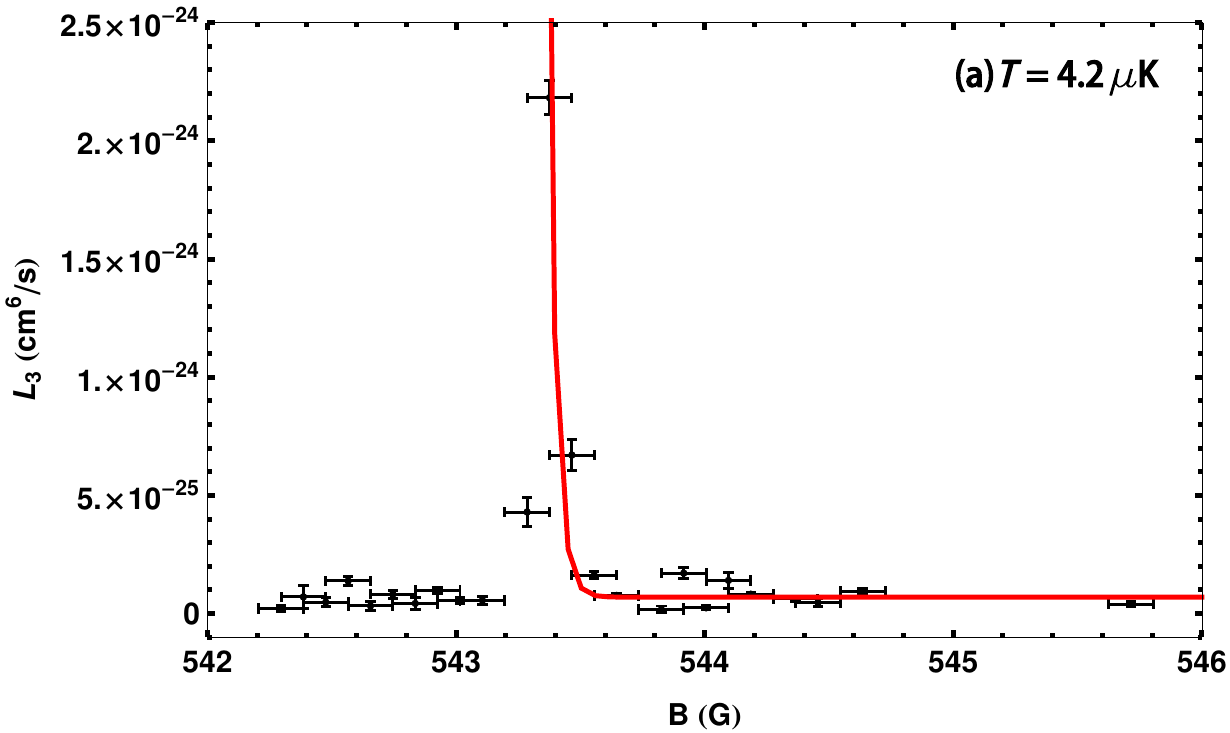} \\
\includegraphics[width=2.5in]{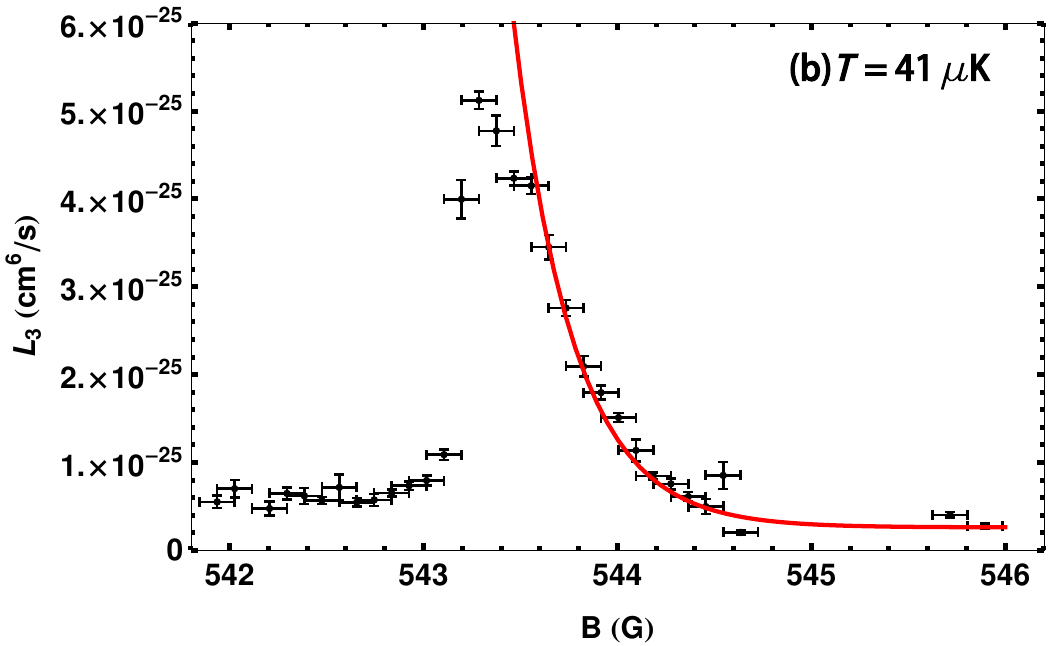}\\
\includegraphics[width=2.5in]{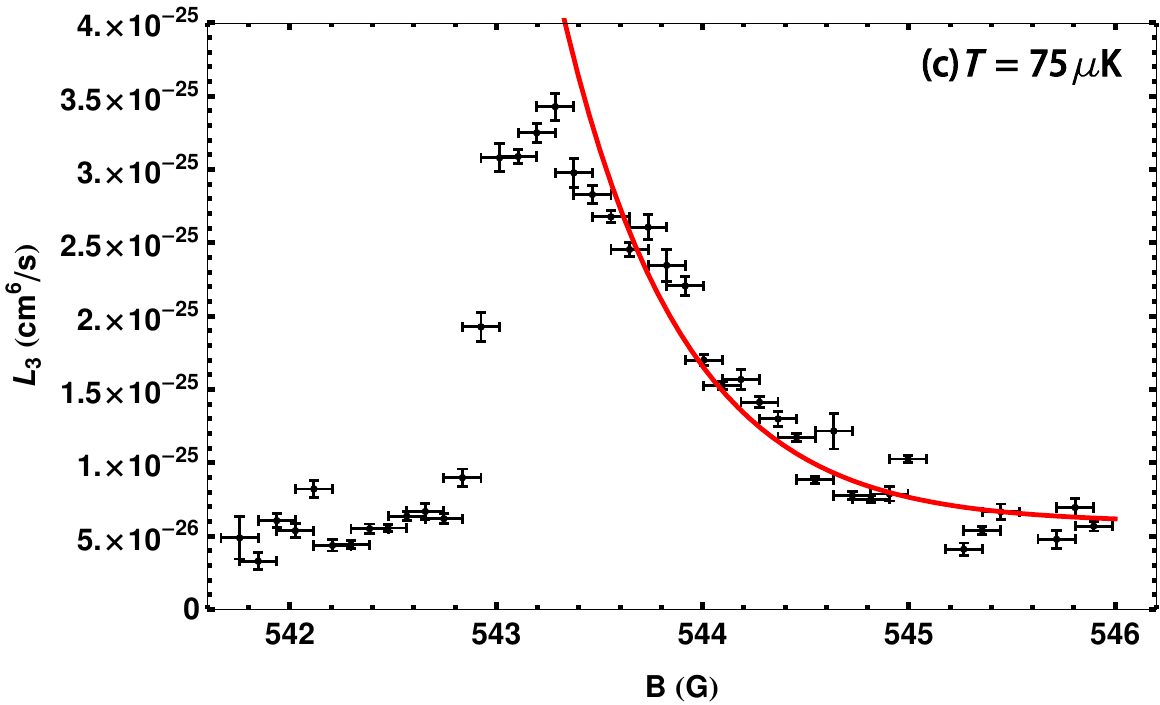}\\
\includegraphics[width=2.5in]{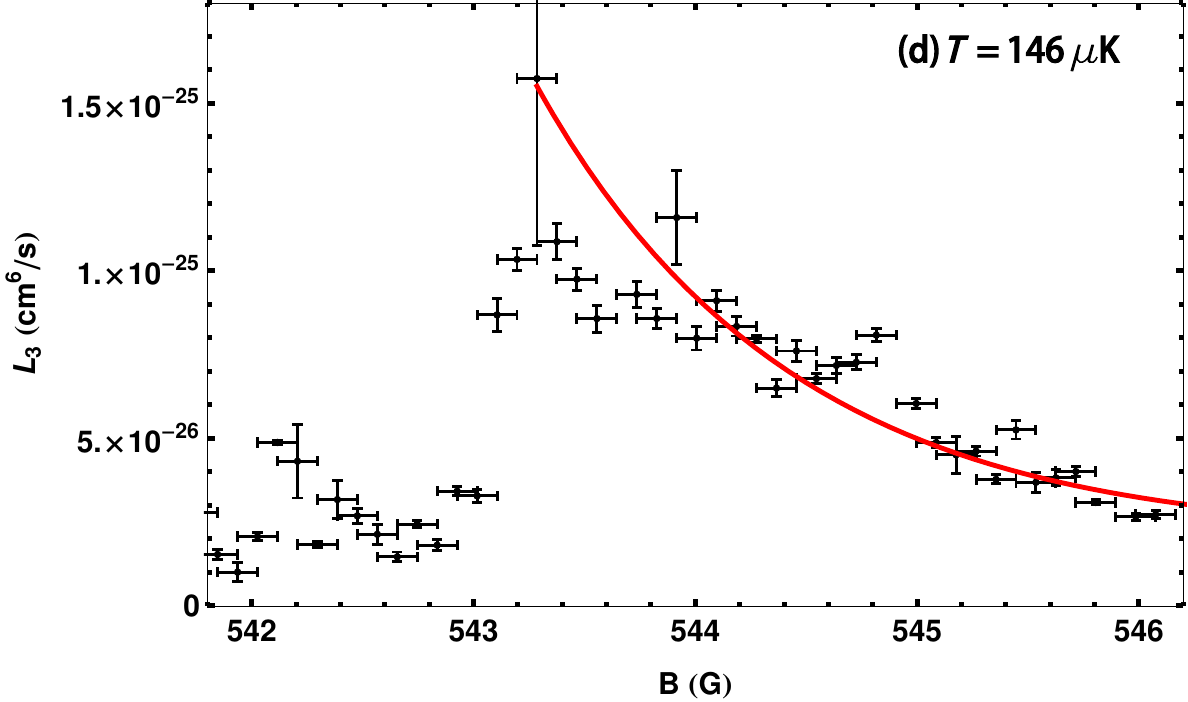}\\
\includegraphics[width=2.5in]{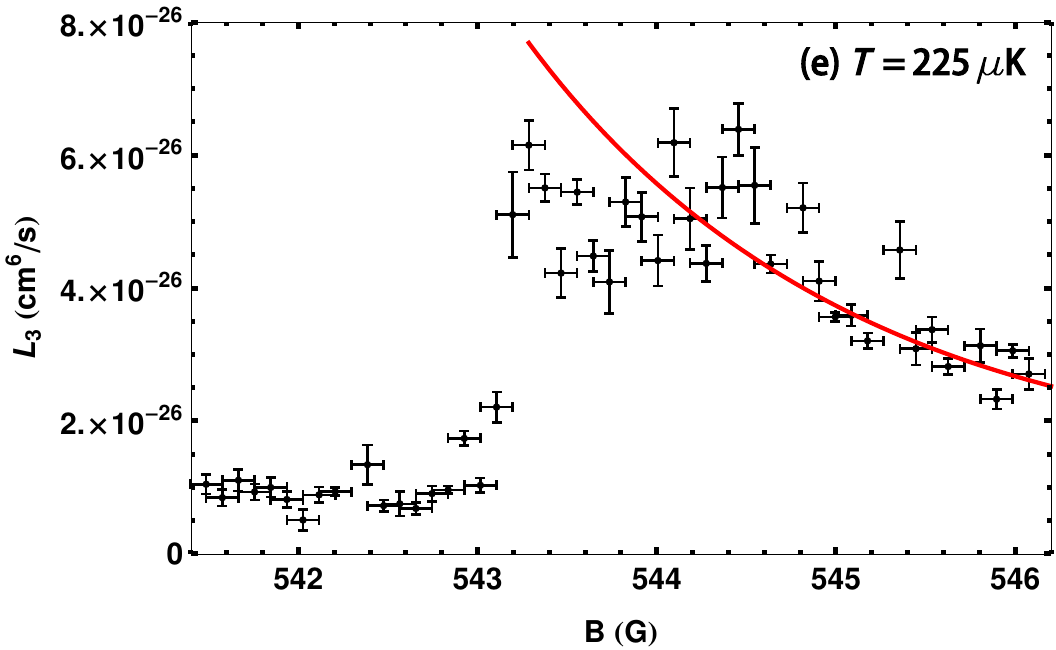}\\
\end{array}$
\end{center}
\caption{$L_{3}(T,B)$ as a function of the magnetic field $B$ at temperatures 4.2$\mu$K (a),\,41$\mu$K (b),\,75$\mu$K (c),\,146$\mu$K (d),\,225$\mu$K (e). The red lines are the fits to theory to be discussed later. The magnetic FR crosses the threshold at $B_0 = 543.25$ G. The trap losses for $B<B_0$ are due to higher-order processes that are not considered in this work.}
\label{fig:atomlossB}
\end{figure}

We measure $L_3$ as a function of the magnetic field at various temperatures from several $\mu$K to 225 $\mu$K. The results are shown in Fig.~\ref{fig:atomlossB}, and will be compared with theory. The highest temperature available is limited by the optical trap depth, where very high trap depth will result in additional heating and loss. For the highest trap depth used in our experiments, we have at least 15 second lifetime for a weak-interacting Fermi gas at 528 G.

\textit{Theory:} Our theory describes three-body recombination via a narrow resonance as an indirect, successive pairwise process. A narrow resonance can be treated as a bound molecular state weakly coupled to a continuum. The time evolution of atomic number densities, $n_a$ and $n_b$ for atoms in states $a$ and $b$ respectively, and the number density $n_{ab}$ of metastable molecules in the resonance state $(ab)_r$, are describe by a set of rate equations
\begin{subequations}
\begin{align}
\frac{dn_a}{dt} =& +(\Gamma_r/\hbar)n_{ab}-K_{ab}n_a n_b \nonumber\\
    & -K^M_{AD}n_a n_{ab} + K^{Br}_{AD}n_a n_{ab}
    +K^{Br}_{AD}n_b n_{ab}\;,\\
\frac{dn_b}{dt} =& +(\Gamma_r/\hbar)n_{ab}-K_{ab}n_a n_b \nonumber\\
    & -K^M_{AD}n_b n_{ab} + K^{Br}_{AD}n_b n_{ab}
    + K^{Br}_{AD}n_a n_{ab}\;,\\
\frac{dn_{ab}}{dt} =& -(\Gamma_r/\hbar)n_{ab}+K_{ab}n_a n_b \nonumber\\
    & -K_{AD}n_a n_{ab} - K_{AD}n_b n_{ab}\;.
\end{align}
\label{eq:rate}
\end{subequations}

Here $\Gamma_r$ is the width of the resonance. $K_{ab}$ is the rate for the formation of metastable molecules via two-body collision at temperature $T$. It is related to the resonance width $\Gamma_r$ by
\begin{equation}
K_{ab} = (\Gamma_r/\hbar)(\sqrt{2}\lambda_T)^3(2l_r+1)e^{-\epsilon_r/k_BT} \;,
\end{equation}
for a resonance in partial wave $l_r$ located at energy $\epsilon_r$. Here $\lambda_T := (2\pi\hbar^2/m k_B T)^{1/2}$ is the thermal wave length of an atom at temperature $T$. The $K^M_{AD}$ in Eq.~(\ref{eq:rate}) is the rate of atom-dimer interaction leading to the formation of a stable molecule, namely for the processes of $a+(ab)_r\to a + (ab)_M$ and $a+(ab)_r\to b + (aa)_M$, or  $b+(ab)_r\to b+(ab)_M$ and $b+(ab)_r\to a + (bb)_M$. $K^{Br}_{AD}$ is the rate of breakup of a metastable molecule via $a+(ab)_r\to 2a + b$ or $b+(ab)_r\to a+2b$. $K_{AD} = K^M_{AD}+K^{Br}_{AD}$ is the total inelastic and reactive rate for atom-dimer interaction. This rate equation ignores the contribution from the direct three-body process to focus on the contribution from the indirect process, which will be shown later to dominate at cold temperatures.

The seemingly complex rate equation, Eq.~(\ref{eq:rate}), simplifies if the $\Gamma_r$ and the measurement time allow $n_{ab}$ to reach a steady state, characterized by $dn_{ab}/dt=0$. In the steady state, one obtains
\begin{equation}
n_{ab} = \frac{(\Gamma_r/\hbar)}{(\Gamma_r/\hbar)+K_{AD}(n_a+n_b)}2^{3/2}\lambda_T^3(2l_r+1)e^{-\epsilon_r/k_BT} \;.
\end{equation}
Under the further initial condition of $n_a(t=0)=n_b(t=0)$, corresponding to our experiment, and the condition of $\Gamma_r/\hbar\gg K_{AD}(n_a+n_b)$, we obtain in steady state $n_a(t)=n_b(t) =: n(t)$ and satisfies Eq.~(\ref{eq:L3def}) with $L_3$ given by
\begin{equation}
L_3(T,\epsilon_r)=3K^M_{AD}(T,\epsilon_r)(\sqrt{2}\lambda_T)^3(2l_r+1)
    e^{-\epsilon_r/k_B T} \;.
\label{eq:L3}
\end{equation}
All the required conditions are well satisfied in our particular experiment. We caution, however, that the typical three-body rate equation, Eq.~(\ref{eq:L3def}), should not be taken for granted for indirect processes. They can have other behaviors under different conditions.

Through Eq.~(\ref{eq:L3}), the rate equation, Eq.~(\ref{eq:rate}), reduces the understanding of our $L_3$ to the understanding of $K^M_{AD}$ which is the rate for the formation of bound molecules in atom interaction with a metastable dimer. This bimolecular process differs from the typical atom-(truly bound) dimer interaction in that its inelastic component does not always leads to the formations of bound molecules even in the limit of zero atom-dimer energy. It can also lead to the breakup of the metastable dimer, resulting in three free atoms. Our theory for $K^M_{AD}$ is based on the multichannel quantum defect theory (MQDT) for reactions and inelastic processes as outlined in Ref.~\cite{Gao10b}. Following an analysis similar to what led to the quantum Langevin (QL) model for reactions \cite{Gao10b,Gao11a}, we obtain
\begin{multline}
K^M_{AD}(T,\epsilon_r) = s^{AD}_K \frac{2}{\sqrt{\pi}}
    \int_0^\infty dx x^{1/2}e^{-x} \\
    \times\sum_{l=0}^\infty \mathcal{M}_{l_rl}(\epsilon_r+k_BTx)
    {\cal W}^{(6)}_{\mathrm{ur}l}(T_s x) \;.
\label{eq:Kmad}
\end{multline}
Here $s^{AD}_K$ is the rate scale for atom-dimer interaction with a van der Waals $-C_6^{AD}/R^6$ long range potential. More specifically, $s^{AD}_K := \pi\hbar\beta_6^{AD}/\mu^{AD}$, where $\mu^{AD}$ is atom-dimer reduced mass and $\beta_6^{AD}:=(2\mu^{AD}C_6^{AD}/\hbar^2)^{1/4}$ is the length scale associated with the atom-dimer van der Waals potential. $\mathcal{M}_{l_rl}(\epsilon_f=\epsilon_r+\epsilon_{AD}):=\sum_{f\in M}|(S^c_{\textrm{eff}})_{fi}|^2$ is a short-range branching ratio for transitions into bound molecular states characterized by set $\{M\}$, with $S^c_{\textrm{eff}}$ being the effective short-range $S$ matrix characterizing atom-dimer, namely three-body interaction within the range of van der Waals length scale \cite{Gao08a,Gao10b}. ${\cal W}^{(6)}_{\mathrm{ur}l}(\epsilon_s)$ is the universal partial inelastic and reactive QL rate for partial wave $l$ \cite{Gao10b}. $T_s := T/s^{AD}_T$ is a scaled temperature, with $s^{AD}_T$ being the temperature scale give by $s^{AD}_T:=s^{AD}_E/k_B$ where $s^{AD}_E:=(\hbar^2/2\mu^{AD})(1/\beta_6^{AD})^2$ is the energy scale associated with $\beta_6^{AD}$.

Equations~(\ref{eq:L3}) and (\ref{eq:Kmad}) provide a foundation for understanding the universal behaviors of three-body recombination via a narrow resonance over a wide range of energies and temperatures. We focus here on a $s$ wave resonance ($l_r=0$) and on the ultracold temperature regime of $T\ll s^{AD}_T$ (namely $T_s\ll 1$), to derive an analytic formula for $L_3$ that is most useful in current experiments. Using the unitarity of an $S$ matrix, we can also write $\mathcal{M}_{l_rl}(\epsilon_f)=1-\sum_{f\in Br}|(S^c_{\textrm{eff}})_{fi}|^2$, namely in terms of the short-range branching ratio into the 3-body breakup channels $\{Br\}$. Taking advantage of the short-range $S$ matrix being insensitive to energy and angular momenta \cite{Gao01,Gao08a}, the short-range branching ratio to bound molecular states is approximately a constant $\mathcal{M}_{l_r=0l=0}(\epsilon_f)\approx \mathscr{M}$ with $\mathscr{M}$ being a dimensionless 3-body parameter related to $S^c_{\textrm{eff}}$ and constrained by $0<\mathscr{M}\le 1$. Substituting this result into Eq.~(\ref{eq:Kmad}), we obtain for an $s$ wave resonance ($l_r=0$) in the ultracold region of $T\ll s^{AD}_T$
\begin{equation}
K^M_{AD}(T,\epsilon_r) \approx \mathscr{M} K^{QL(6)}_{AD}(T_s) \;,
\label{eq:Kmads}
\end{equation}
with $K^{QL(6)}_{AD}(T_s)$ being the universal QL rate that is well approximated in the ultracold $s$ wave region by \cite{Gao10b}
\begin{equation}
K^{QL(6)}_{AD}(T_s) \approx s^{AD}_K 4\bar{a}^{(6)}_{sl=0}\left(1
    -\frac{4\bar{a}^{(6)}_{sl=0}}{\sqrt{\pi}}T_s^{1/2}\right)\;,
\label{eq:KQL6}
\end{equation}
where $\bar{a}^{(6)}_{sl=0} = 2\pi/[\Gamma(1/4)]^{2} \approx 0.4779888$ is a universal number that represents the scaled mean $s$ wave scattering length for a $-1/R^6$-type van der Waals potential \cite{Gao09a}.

Substituting Eq.~(\ref{eq:Kmads}) into Eq.~(\ref{eq:L3}), we obtain
\begin{equation}
L_3(T,\epsilon_r)\approx 3\mathscr{M} K^{QL(6)}_{AD}(T_s)(\sqrt{2}\lambda_T)^3 e^{-\epsilon_r/k_B T} \;.
\label{eq:L3s}
\end{equation}
In the presence of a narrow $s$ wave resonance within the ultracold energy regime above the threshold, Equation~(\ref{eq:L3s}) gives an analytic description of the three-body recombination rate $L_3$ as a function of both the temperature and the resonance position, in terms of a single dimensionless three-body parameter $0<\mathscr{M}\le 1$. The resonance can in principle be of any origin, but a magnetic FR offers a unique opportunity to tune the resonance position, and thus to test the predicted dependence on $\epsilon_r$.

\begin{figure}
\includegraphics[width=\columnwidth]{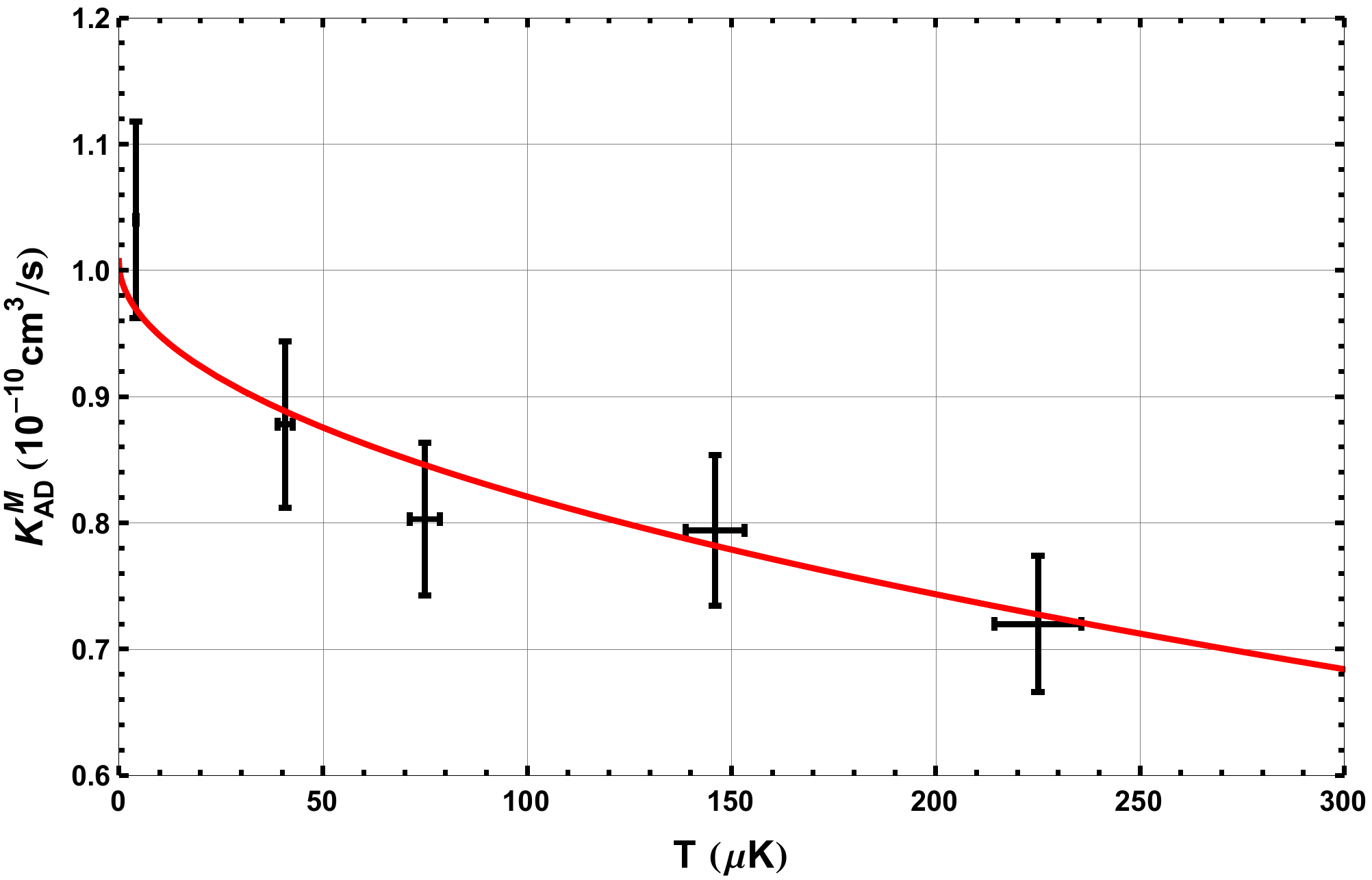}
\caption{The rate constant $K^M_{AD}(T)$ near the $^6$Li narrow $s$-wave FR resonance at 543.25 G. The red solid line is a fit of our theoretical model, Eqs.~(\ref{eq:Kmads}) and (\ref{eq:KQL6}), to our experimental measurements from which the three-body parameter $\mathscr{M}= 0.25\pm0.01$ is extracted.}
\label{fig:Kad}
\end{figure}

\begin{table}
\begin{ruledtabular}
\begin{tabular}{rrrr}
  $T$ ($\mu$K) & $\delta T$ ($\mu$K) & $K^M_{AD}$ ($10^{-10}$ cm$^3$/s) & $\delta K^M_{AD}$ ($10^{-12}$ cm$^3$/s) \\
  \hline
    4.2 & 0.2 & $1.04 $ & $8$  \\
    41  & 4   & $0.878$ & $7$ \\
   75   & 4   & $0.803$ & $6$ \\
   146  & 7   & $0.794$ & $6$ \\
  225   & 11  & $0.720$ & $5$ \\
\end{tabular}
\end{ruledtabular}
\caption{The measured results and error bars of $K^M_{AD}(T)$ }
\label{tab:Kadtable}
\end{table}

\textit{Comparison between theory and experiment:} For $^6$Li, using $C_6 = 1393.39$ a.u. \cite{Yan96} for the atom-atom potential, we have $C_6^{AD} \approx 2C_6= 2786.78$ a.u., from which we have $\beta_6^{AD} = 79.8935$ a.u., $s^{AD}_T = 3383.85$ $\mu$K, and $s^{AD}_K = 2.1035\times 10^{-10}$ cm$^3$/s. For our particular Feshbach resonance, the resonance position is given by $\epsilon_r =\mu_r(B-B_{0})$ with $\mu_r= 1.98 \mu_B$ being the differential magnetic moment for the resonance \cite{Chin10}. Equation~(\ref{eq:L3}) now gives us
\begin{equation}
L_3(T, B)\approx 3K^M_{AD}(T)(\sqrt{2}\lambda_T)^3
    \exp\left[-\frac{\mu_r(B-B_0)}{k_B T}\right] \;.
\label{eq:L3sB}
\end{equation}
Figure~\ref{fig:atomlossB} shows the fits of this equation to experimental loss spectra, giving experimental results of $K^M_{AD}(T)$ at five different temperatures, tabulated in Table~\ref{tab:Kadtable} and plotted in Fig.~\ref{fig:Kad}. Our result for $K^M_{AD}$ at the lower end of the temperatures, 4.2 $\mu$K, is consistent with the earlier result of Hazlett \textit{et al.} \cite{Hazlett12}. Figure~\ref{fig:Kad} further shows that the temperature dependence of the rate $K^M_{AD}(T)$ is well described by analytic formulas, Eqs.~(\ref{eq:Kmads}) and (\ref{eq:KQL6}), a fit to which gives us the three-body parameter $\mathscr{M}= 0.25\pm0.01$, consistent with $0<\mathscr{M}\le 1$.

\textit{Discussions and conclusions:} We have measured and theoretically analyzed that three-body recombination around a narrow resonance, specifically a narrow $s$ wave resonance. We have shown that it follows a universal behavior determined by the long-range van der Waals potential with a single three-body parameter $\mathscr{M}$. When applied to a magnetic FR, the theory gives the line shape of the Feshbach spectrum, namely $L_3$ vs $B$, described by Eq.~(\ref{eq:L3sB}). It shows that the line shape is temperature-dependent and has a width of the order of $k_B T/\mu_B$ (see also Ref.~\cite{Hazlett12}).

The theory further shows that in the presence of a narrow $s$-wave resonance within $k_B T$ above the threshold, this indirect (resonant) process gives rise to a three-body recombination rate of the order of $(\hbar/m)\beta_6\lambda_T^3$, which, at ultracold temperatures of $T\ll s_T$, is much greater, by a factor of $(\lambda_T/\beta_6)^3$, than that for the direct (background or nonresonant) process, which can be estimated to be of the order of $(\hbar/m)\beta_6^4$. Thus at ultracold temperatures, the indirect process dominate the three-body recombination if there is a narrow $s$-wave resonance within $k_B T$ above the (two-body) threshold.

Many of the concepts of this work are applicable to resonances in nonzero partial waves (see, e.g., Refs.~\cite{Regal03,Ticknor04,Gao17b,Yao17}), the understanding of which will further expand the temperature regime of three-body physics towards practical chemistry \cite{Forrey13,Greene14,Forrey15,Wei97,Pack98,Suno02}. More measurements of $\mathscr{M}$ for other narrow resonances and other systems will further stimulate a deeper understanding of this three-body parameter. It can be expected to be related in a universal manner to short-range $K^c$ matrix parameters $K^c_S$ and $K^c_T$ for atom-atom interaction in (electronic) spin singlet and triplet, respectively \cite{Gao05a}. Such a relationship, when revealed and understood, would signal the arrival of a QDT for few-atom systems, and will represent a big step forward in few-body physics and in chemistry.

\begin{acknowledgments}
Le Luo is a member of the Indiana University Center for Spacetime Symmetries (IUCSS). Le Luo thanks supports from Indiana University IUCRG, RSFG, Purdue University PRF, CNSF-11774436. Bo Gao is supported by NSF under grant No. PHY-1607256.
\end{acknowledgments}


\begin{thebibliography}{41}%
\makeatletter
\providecommand \@ifxundefined [1]{%
 \@ifx{#1\undefined}
}%
\providecommand \@ifnum [1]{%
 \ifnum #1\expandafter \@firstoftwo
 \else \expandafter \@secondoftwo
 \fi
}%
\providecommand \@ifx [1]{%
 \ifx #1\expandafter \@firstoftwo
 \else \expandafter \@secondoftwo
 \fi
}%
\providecommand \natexlab [1]{#1}%
\providecommand \enquote  [1]{``#1''}%
\providecommand \bibnamefont  [1]{#1}%
\providecommand \bibfnamefont [1]{#1}%
\providecommand \citenamefont [1]{#1}%
\providecommand \href@noop [0]{\@secondoftwo}%
\providecommand \href [0]{\begingroup \@sanitize@url \@href}%
\providecommand \@href[1]{\@@startlink{#1}\@@href}%
\providecommand \@@href[1]{\endgroup#1\@@endlink}%
\providecommand \@sanitize@url [0]{\catcode `\\12\catcode `\$12\catcode
  `\&12\catcode `\#12\catcode `\^12\catcode `\_12\catcode `\%12\relax}%
\providecommand \@@startlink[1]{}%
\providecommand \@@endlink[0]{}%
\providecommand \url  [0]{\begingroup\@sanitize@url \@url }%
\providecommand \@url [1]{\endgroup\@href {#1}{\urlprefix }}%
\providecommand \urlprefix  [0]{URL }%
\providecommand \Eprint [0]{\href }%
\providecommand \doibase [0]{http://dx.doi.org/}%
\providecommand \selectlanguage [0]{\@gobble}%
\providecommand \bibinfo  [0]{\@secondoftwo}%
\providecommand \bibfield  [0]{\@secondoftwo}%
\providecommand \translation [1]{[#1]}%
\providecommand \BibitemOpen [0]{}%
\providecommand \bibitemStop [0]{}%
\providecommand \bibitemNoStop [0]{.\EOS\space}%
\providecommand \EOS [0]{\spacefactor3000\relax}%
\providecommand \BibitemShut  [1]{\csname bibitem#1\endcsname}%
\let\auto@bib@innerbib\@empty
\bibitem [{\citenamefont {Turk}\ \emph {et~al.}(2011)\citenamefont {Turk},
  \citenamefont {Clark}, \citenamefont {Glover}, \citenamefont {Greif},
  \citenamefont {Abel}, \citenamefont {Klessen},\ and\ \citenamefont
  {Bromm}}]{Turk11}%
  \BibitemOpen
  \bibfield  {author} {\bibinfo {author} {\bibfnamefont {M.~J.}\ \bibnamefont
  {Turk}}, \bibinfo {author} {\bibfnamefont {P.}~\bibnamefont {Clark}},
  \bibinfo {author} {\bibfnamefont {S.~C.~O.}\ \bibnamefont {Glover}}, \bibinfo
  {author} {\bibfnamefont {T.~H.}\ \bibnamefont {Greif}}, \bibinfo {author}
  {\bibfnamefont {T.}~\bibnamefont {Abel}}, \bibinfo {author} {\bibfnamefont
  {R.}~\bibnamefont {Klessen}}, \ and\ \bibinfo {author} {\bibfnamefont
  {V.}~\bibnamefont {Bromm}},\ }\href
  {http://stacks.iop.org/0004-637X/726/i=1/a=55} {\bibfield  {journal}
  {\bibinfo  {journal} {The Astrophysical Journal}\ }\textbf {\bibinfo {volume}
  {726}},\ \bibinfo {pages} {55} (\bibinfo {year} {2011})}\BibitemShut
  {NoStop}%
\bibitem [{\citenamefont {Forrey}(2013)}]{Forrey13}%
  \BibitemOpen
  \bibfield  {author} {\bibinfo {author} {\bibfnamefont {R.~C.}\ \bibnamefont
  {Forrey}},\ }\href {http://stacks.iop.org/2041-8205/773/i=2/a=L25} {\bibfield
   {journal} {\bibinfo  {journal} {The Astrophysical Journal Letters}\ }\textbf
  {\bibinfo {volume} {773}},\ \bibinfo {pages} {L25} (\bibinfo {year}
  {2013})}\BibitemShut {NoStop}%
\bibitem [{\citenamefont {Pérez-Ríos}\ \emph {et~al.}(2014)\citenamefont
  {Pérez-Ríos}, \citenamefont {Ragole}, \citenamefont {Wang},\ and\
  \citenamefont {Greene}}]{Greene14}%
  \BibitemOpen
  \bibfield  {author} {\bibinfo {author} {\bibfnamefont {J.}~\bibnamefont
  {Pérez-Ríos}}, \bibinfo {author} {\bibfnamefont {S.}~\bibnamefont
  {Ragole}}, \bibinfo {author} {\bibfnamefont {J.}~\bibnamefont {Wang}}, \ and\
  \bibinfo {author} {\bibfnamefont {C.~H.}\ \bibnamefont {Greene}},\ }\href
  {http://scitation.aip.org/content/aip/journal/jcp/140/4/10.1063/1.4861851}
  {\bibfield  {journal} {\bibinfo  {journal} {The Journal of Chemical Physics}\
  }\textbf {\bibinfo {volume} {140}},\ \bibinfo {eid} {044307} (\bibinfo {year}
  {2014})}\BibitemShut {NoStop}%
\bibitem [{\citenamefont {Forrey}(2015)}]{Forrey15}%
  \BibitemOpen
  \bibfield  {author} {\bibinfo {author} {\bibfnamefont {R.~C.}\ \bibnamefont
  {Forrey}},\ }\href
  {http://scitation.aip.org/content/aip/journal/jcp/143/2/10.1063/1.4926325}
  {\bibfield  {journal} {\bibinfo  {journal} {The Journal of Chemical Physics}\
  }\textbf {\bibinfo {volume} {143}},\ \bibinfo {eid} {024101} (\bibinfo {year}
  {2015})}\BibitemShut {NoStop}%
\bibitem [{\citenamefont {Wei}\ \emph {et~al.}(1997)\citenamefont {Wei},
  \citenamefont {Alavi},\ and\ \citenamefont {Snider}}]{Wei97}%
  \BibitemOpen
  \bibfield  {author} {\bibinfo {author} {\bibfnamefont {G.~W.}\ \bibnamefont
  {Wei}}, \bibinfo {author} {\bibfnamefont {S.}~\bibnamefont {Alavi}}, \ and\
  \bibinfo {author} {\bibfnamefont {R.~F.}\ \bibnamefont {Snider}},\ }\href
  {\doibase http://dx.doi.org/10.1063/1.473294} {\bibfield  {journal} {\bibinfo
   {journal} {The Journal of Chemical Physics}\ }\textbf {\bibinfo {volume}
  {106}},\ \bibinfo {pages} {1463} (\bibinfo {year} {1997})}\BibitemShut
  {NoStop}%
\bibitem [{\citenamefont {T~Pack}\ \emph {et~al.}(1998)\citenamefont {T~Pack},
  \citenamefont {Walker},\ and\ \citenamefont {Kendrick}}]{Pack98}%
  \BibitemOpen
  \bibfield  {author} {\bibinfo {author} {\bibfnamefont {R.}~\bibnamefont
  {T~Pack}}, \bibinfo {author} {\bibfnamefont {R.~B.}\ \bibnamefont {Walker}},
  \ and\ \bibinfo {author} {\bibfnamefont {B.~K.}\ \bibnamefont {Kendrick}},\
  }\href {\doibase http://dx.doi.org/10.1063/1.477348} {\bibfield  {journal}
  {\bibinfo  {journal} {The Journal of Chemical Physics}\ }\textbf {\bibinfo
  {volume} {109}},\ \bibinfo {pages} {6701} (\bibinfo {year}
  {1998})}\BibitemShut {NoStop}%
\bibitem [{\citenamefont {Suzuki}\ and\ \citenamefont
  {Varga}(1998)}]{Suzuki98}%
  \BibitemOpen
  \bibfield  {author} {\bibinfo {author} {\bibfnamefont {Y.}~\bibnamefont
  {Suzuki}}\ and\ \bibinfo {author} {\bibfnamefont {K.}~\bibnamefont {Varga}},\
  }\href@noop {} {\emph {\bibinfo {title} {Stochastic Variational Approach to
  Quantum-Mechanical Few-Body Problems}}}\ (\bibinfo  {publisher}
  {Springer-Verlag, Berlin},\ \bibinfo {year} {1998})\BibitemShut {NoStop}%
\bibitem [{\citenamefont {Suno}\ \emph {et~al.}(2002)\citenamefont {Suno},
  \citenamefont {Esry}, \citenamefont {Greene},\ and\ \citenamefont {{Burke,
  Jr.}}}]{Suno02}%
  \BibitemOpen
  \bibfield  {author} {\bibinfo {author} {\bibfnamefont {H.}~\bibnamefont
  {Suno}}, \bibinfo {author} {\bibfnamefont {B.~D.}\ \bibnamefont {Esry}},
  \bibinfo {author} {\bibfnamefont {C.~H.}\ \bibnamefont {Greene}}, \ and\
  \bibinfo {author} {\bibfnamefont {J.~P.}\ \bibnamefont {{Burke, Jr.}}},\
  }\href@noop {} {\bibfield  {journal} {\bibinfo  {journal} {Phys. Rev. A}\
  }\textbf {\bibinfo {volume} {65}},\ \bibinfo {pages} {042725} (\bibinfo
  {year} {2002})}\BibitemShut {NoStop}%
\bibitem [{\citenamefont {Chin}\ \emph {et~al.}(2010)\citenamefont {Chin},
  \citenamefont {Grimm}, \citenamefont {Julienne},\ and\ \citenamefont
  {Tiesinga}}]{Chin10}%
  \BibitemOpen
  \bibfield  {author} {\bibinfo {author} {\bibfnamefont {C.}~\bibnamefont
  {Chin}}, \bibinfo {author} {\bibfnamefont {R.}~\bibnamefont {Grimm}},
  \bibinfo {author} {\bibfnamefont {P.}~\bibnamefont {Julienne}}, \ and\
  \bibinfo {author} {\bibfnamefont {E.}~\bibnamefont {Tiesinga}},\ }\href
  {\doibase 10.1103/RevModPhys.82.1225} {\bibfield  {journal} {\bibinfo
  {journal} {Rev. Mod. Phys.}\ }\textbf {\bibinfo {volume} {82}},\ \bibinfo
  {pages} {1225} (\bibinfo {year} {2010})}\BibitemShut {NoStop}%
\bibitem [{\citenamefont {Nielsen}\ \emph {et~al.}(2001)\citenamefont
  {Nielsen}, \citenamefont {Fedorov}, \citenamefont {Jensen},\ and\
  \citenamefont {Garrido}}]{Nielsen01}%
  \BibitemOpen
  \bibfield  {author} {\bibinfo {author} {\bibfnamefont {E.}~\bibnamefont
  {Nielsen}}, \bibinfo {author} {\bibfnamefont {D.}~\bibnamefont {Fedorov}},
  \bibinfo {author} {\bibfnamefont {A.}~\bibnamefont {Jensen}}, \ and\ \bibinfo
  {author} {\bibfnamefont {E.}~\bibnamefont {Garrido}},\ }\href {\doibase
  https://doi.org/10.1016/S0370-1573(00)00107-1} {\bibfield  {journal}
  {\bibinfo  {journal} {Physics Reports}\ }\textbf {\bibinfo {volume} {347}},\
  \bibinfo {pages} {373 } (\bibinfo {year} {2001})}\BibitemShut {NoStop}%
\bibitem [{\citenamefont {Braaten}\ and\ \citenamefont {Hammer}(2006)}]{BH06}%
  \BibitemOpen
  \bibfield  {author} {\bibinfo {author} {\bibfnamefont {E.}~\bibnamefont
  {Braaten}}\ and\ \bibinfo {author} {\bibfnamefont {H.-W.}\ \bibnamefont
  {Hammer}},\ }\href {\doibase DOI: 10.1016/j.physrep.2006.03.001} {\bibfield
  {journal} {\bibinfo  {journal} {Physics Reports}\ }\textbf {\bibinfo {volume}
  {428}},\ \bibinfo {pages} {259 } (\bibinfo {year} {2006})}\BibitemShut
  {NoStop}%
\bibitem [{\citenamefont {Kraemer}\ \emph {et~al.}(2006)\citenamefont
  {Kraemer}, \citenamefont {Mark}, \citenamefont {Waldburger}, \citenamefont
  {Danzl}, \citenamefont {Chin}, \citenamefont {Engeser}, \citenamefont
  {Lange}, \citenamefont {Pilch}, \citenamefont {Jaakkola}, \citenamefont
  {N\"{a}gerl},\ and\ \citenamefont {Grimm}}]{Kraemer06}%
  \BibitemOpen
  \bibfield  {author} {\bibinfo {author} {\bibfnamefont {T.}~\bibnamefont
  {Kraemer}}, \bibinfo {author} {\bibfnamefont {M.}~\bibnamefont {Mark}},
  \bibinfo {author} {\bibfnamefont {P.}~\bibnamefont {Waldburger}}, \bibinfo
  {author} {\bibfnamefont {J.~G.}\ \bibnamefont {Danzl}}, \bibinfo {author}
  {\bibfnamefont {C.}~\bibnamefont {Chin}}, \bibinfo {author} {\bibfnamefont
  {B.}~\bibnamefont {Engeser}}, \bibinfo {author} {\bibfnamefont {A.~D.}\
  \bibnamefont {Lange}}, \bibinfo {author} {\bibfnamefont {K.}~\bibnamefont
  {Pilch}}, \bibinfo {author} {\bibfnamefont {A.}~\bibnamefont {Jaakkola}},
  \bibinfo {author} {\bibfnamefont {H.-C.}\ \bibnamefont {N\"{a}gerl}}, \ and\
  \bibinfo {author} {\bibfnamefont {R.}~\bibnamefont {Grimm}},\ }\href
  {\doibase doi:10.1038/nature04626} {\bibfield  {journal} {\bibinfo  {journal}
  {Nature}\ }\textbf {\bibinfo {volume} {440}},\ \bibinfo {pages} {315}
  (\bibinfo {year} {2006})}\BibitemShut {NoStop}%
\bibitem [{\citenamefont {Greene}(2010)}]{Greene10}%
  \BibitemOpen
  \bibfield  {author} {\bibinfo {author} {\bibfnamefont {C.~H.}\ \bibnamefont
  {Greene}},\ }\href@noop {} {\bibfield  {journal} {\bibinfo  {journal}
  {Physics Today}\ }\textbf {\bibinfo {volume} {63}},\ \bibinfo {pages} {40}
  (\bibinfo {year} {2010})}\BibitemShut {NoStop}%
\bibitem [{\citenamefont {Berninger}\ \emph {et~al.}(2011)\citenamefont
  {Berninger}, \citenamefont {Zenesini}, \citenamefont {Huang}, \citenamefont
  {Harm}, \citenamefont {N\"agerl}, \citenamefont {Ferlaino}, \citenamefont
  {Grimm}, \citenamefont {Julienne},\ and\ \citenamefont
  {Hutson}}]{Berninger11}%
  \BibitemOpen
  \bibfield  {author} {\bibinfo {author} {\bibfnamefont {M.}~\bibnamefont
  {Berninger}}, \bibinfo {author} {\bibfnamefont {A.}~\bibnamefont {Zenesini}},
  \bibinfo {author} {\bibfnamefont {B.}~\bibnamefont {Huang}}, \bibinfo
  {author} {\bibfnamefont {W.}~\bibnamefont {Harm}}, \bibinfo {author}
  {\bibfnamefont {H.-C.}\ \bibnamefont {N\"agerl}}, \bibinfo {author}
  {\bibfnamefont {F.}~\bibnamefont {Ferlaino}}, \bibinfo {author}
  {\bibfnamefont {R.}~\bibnamefont {Grimm}}, \bibinfo {author} {\bibfnamefont
  {P.~S.}\ \bibnamefont {Julienne}}, \ and\ \bibinfo {author} {\bibfnamefont
  {J.~M.}\ \bibnamefont {Hutson}},\ }\href {\doibase
  10.1103/PhysRevLett.107.120401} {\bibfield  {journal} {\bibinfo  {journal}
  {Phys. Rev. Lett.}\ }\textbf {\bibinfo {volume} {107}},\ \bibinfo {pages}
  {120401} (\bibinfo {year} {2011})}\BibitemShut {NoStop}%
\bibitem [{\citenamefont {Wang}\ \emph
  {et~al.}(2012{\natexlab{a}})\citenamefont {Wang}, \citenamefont {D'Incao},
  \citenamefont {Esry},\ and\ \citenamefont {Greene}}]{Wang12}%
  \BibitemOpen
  \bibfield  {author} {\bibinfo {author} {\bibfnamefont {J.}~\bibnamefont
  {Wang}}, \bibinfo {author} {\bibfnamefont {J.~P.}\ \bibnamefont {D'Incao}},
  \bibinfo {author} {\bibfnamefont {B.~D.}\ \bibnamefont {Esry}}, \ and\
  \bibinfo {author} {\bibfnamefont {C.~H.}\ \bibnamefont {Greene}},\ }\href
  {\doibase 10.1103/PhysRevLett.108.263001} {\bibfield  {journal} {\bibinfo
  {journal} {Phys. Rev. Lett.}\ }\textbf {\bibinfo {volume} {108}},\ \bibinfo
  {pages} {263001} (\bibinfo {year} {2012}{\natexlab{a}})}\BibitemShut
  {NoStop}%
\bibitem [{\citenamefont {Wang}\ \emph
  {et~al.}(2012{\natexlab{b}})\citenamefont {Wang}, \citenamefont {Wang},
  \citenamefont {D'Incao},\ and\ \citenamefont {Greene}}]{Wang12b}%
  \BibitemOpen
  \bibfield  {author} {\bibinfo {author} {\bibfnamefont {Y.}~\bibnamefont
  {Wang}}, \bibinfo {author} {\bibfnamefont {J.}~\bibnamefont {Wang}}, \bibinfo
  {author} {\bibfnamefont {J.~P.}\ \bibnamefont {D'Incao}}, \ and\ \bibinfo
  {author} {\bibfnamefont {C.~H.}\ \bibnamefont {Greene}},\ }\href {\doibase
  10.1103/PhysRevLett.109.243201} {\bibfield  {journal} {\bibinfo  {journal}
  {Phys. Rev. Lett.}\ }\textbf {\bibinfo {volume} {109}},\ \bibinfo {pages}
  {243201} (\bibinfo {year} {2012}{\natexlab{b}})}\BibitemShut {NoStop}%
\bibitem [{\citenamefont {Naidon}\ \emph {et~al.}(2014)\citenamefont {Naidon},
  \citenamefont {Endo},\ and\ \citenamefont {Ueda}}]{Naidon14}%
  \BibitemOpen
  \bibfield  {author} {\bibinfo {author} {\bibfnamefont {P.}~\bibnamefont
  {Naidon}}, \bibinfo {author} {\bibfnamefont {S.}~\bibnamefont {Endo}}, \ and\
  \bibinfo {author} {\bibfnamefont {M.}~\bibnamefont {Ueda}},\ }\href {\doibase
  10.1103/PhysRevLett.112.105301} {\bibfield  {journal} {\bibinfo  {journal}
  {Phys. Rev. Lett.}\ }\textbf {\bibinfo {volume} {112}},\ \bibinfo {pages}
  {105301} (\bibinfo {year} {2014})}\BibitemShut {NoStop}%
\bibitem [{\citenamefont {Blume}(2015)}]{Blume15}%
  \BibitemOpen
  \bibfield  {author} {\bibinfo {author} {\bibfnamefont {D.}~\bibnamefont
  {Blume}},\ }\href {\doibase 10.1007/s00601-015-0996-6} {\bibfield  {journal}
  {\bibinfo  {journal} {Few-Body Systems}\ }\textbf {\bibinfo {volume} {56}},\
  \bibinfo {pages} {859} (\bibinfo {year} {2015})}\BibitemShut {NoStop}%
\bibitem [{\citenamefont {Mestrom}\ \emph {et~al.}(2017)\citenamefont
  {Mestrom}, \citenamefont {Wang}, \citenamefont {Greene},\ and\ \citenamefont
  {D'Incao}}]{Mestrom17}%
  \BibitemOpen
  \bibfield  {author} {\bibinfo {author} {\bibfnamefont {P.~M.~A.}\
  \bibnamefont {Mestrom}}, \bibinfo {author} {\bibfnamefont {J.}~\bibnamefont
  {Wang}}, \bibinfo {author} {\bibfnamefont {C.~H.}\ \bibnamefont {Greene}}, \
  and\ \bibinfo {author} {\bibfnamefont {J.~P.}\ \bibnamefont {D'Incao}},\
  }\href {\doibase 10.1103/PhysRevA.95.032707} {\bibfield  {journal} {\bibinfo
  {journal} {Phys. Rev. A}\ }\textbf {\bibinfo {volume} {95}},\ \bibinfo
  {pages} {032707} (\bibinfo {year} {2017})}\BibitemShut {NoStop}%
\bibitem [{\citenamefont {Johansen}\ \emph {et~al.}(2017)\citenamefont
  {Johansen}, \citenamefont {DeSalvo}, \citenamefont {Patel},\ and\
  \citenamefont {Chin}}]{Johansen17}%
  \BibitemOpen
  \bibfield  {author} {\bibinfo {author} {\bibfnamefont {J.}~\bibnamefont
  {Johansen}}, \bibinfo {author} {\bibfnamefont {B.~J.}\ \bibnamefont
  {DeSalvo}}, \bibinfo {author} {\bibfnamefont {K.}~\bibnamefont {Patel}}, \
  and\ \bibinfo {author} {\bibfnamefont {C.}~\bibnamefont {Chin}},\ }\href
  {\doibase 10.1038/nphys4130} {\bibfield  {journal} {\bibinfo  {journal}
  {Nature Physics}\ }\textbf {\bibinfo {volume} {13}},\ \bibinfo {pages} {731}
  (\bibinfo {year} {2017})}\BibitemShut {NoStop}%
\bibitem [{\citenamefont {Hazlett}\ \emph {et~al.}(2012)\citenamefont
  {Hazlett}, \citenamefont {Zhang}, \citenamefont {Stites},\ and\ \citenamefont
  {O'Hara}}]{Hazlett12}%
  \BibitemOpen
  \bibfield  {author} {\bibinfo {author} {\bibfnamefont {E.~L.}\ \bibnamefont
  {Hazlett}}, \bibinfo {author} {\bibfnamefont {Y.}~\bibnamefont {Zhang}},
  \bibinfo {author} {\bibfnamefont {R.~W.}\ \bibnamefont {Stites}}, \ and\
  \bibinfo {author} {\bibfnamefont {K.~M.}\ \bibnamefont {O'Hara}},\ }\href
  {\doibase 10.1103/PhysRevLett.108.045304} {\bibfield  {journal} {\bibinfo
  {journal} {Phys. Rev. Lett.}\ }\textbf {\bibinfo {volume} {108}},\ \bibinfo
  {pages} {045304} (\bibinfo {year} {2012})}\BibitemShut {NoStop}%
\bibitem [{\citenamefont {Wang}\ \emph {et~al.}(2013)\citenamefont {Wang},
  \citenamefont {Heo}, \citenamefont {Rvachov}, \citenamefont {Cotta},\ and\
  \citenamefont {Ketterle}}]{Wang13}%
  \BibitemOpen
  \bibfield  {author} {\bibinfo {author} {\bibfnamefont {T.~T.}\ \bibnamefont
  {Wang}}, \bibinfo {author} {\bibfnamefont {M.-S.}\ \bibnamefont {Heo}},
  \bibinfo {author} {\bibfnamefont {T.~M.}\ \bibnamefont {Rvachov}}, \bibinfo
  {author} {\bibfnamefont {D.~A.}\ \bibnamefont {Cotta}}, \ and\ \bibinfo
  {author} {\bibfnamefont {W.}~\bibnamefont {Ketterle}},\ }\href {\doibase
  10.1103/PhysRevLett.110.173203} {\bibfield  {journal} {\bibinfo  {journal}
  {Phys. Rev. Lett.}\ }\textbf {\bibinfo {volume} {110}},\ \bibinfo {pages}
  {173203} (\bibinfo {year} {2013})}\BibitemShut {NoStop}%
\bibitem [{\citenamefont {Gao}(2003)}]{Gao03}%
  \BibitemOpen
  \bibfield  {author} {\bibinfo {author} {\bibfnamefont {B.}~\bibnamefont
  {Gao}},\ }\href {http://stacks.iop.org/0953-4075/36/i=10/a=319} {\bibfield
  {journal} {\bibinfo  {journal} {Journal of Physics B: Atomic, Molecular and
  Optical Physics}\ }\textbf {\bibinfo {volume} {36}},\ \bibinfo {pages} {2111}
  (\bibinfo {year} {2003})}\BibitemShut {NoStop}%
\bibitem [{\citenamefont {Gao}(2008)}]{Gao08a}%
  \BibitemOpen
  \bibfield  {author} {\bibinfo {author} {\bibfnamefont {B.}~\bibnamefont
  {Gao}},\ }\href {\doibase 10.1103/PhysRevA.78.012702} {\bibfield  {journal}
  {\bibinfo  {journal} {Phys. Rev. A}\ }\textbf {\bibinfo {volume} {78}},\
  \bibinfo {pages} {012702} (\bibinfo {year} {2008})}\BibitemShut {NoStop}%
\bibitem [{\citenamefont {Gao}\ \emph {et~al.}(2005)\citenamefont {Gao},
  \citenamefont {Tiesinga}, \citenamefont {Williams},\ and\ \citenamefont
  {Julienne}}]{Gao05a}%
  \BibitemOpen
  \bibfield  {author} {\bibinfo {author} {\bibfnamefont {B.}~\bibnamefont
  {Gao}}, \bibinfo {author} {\bibfnamefont {E.}~\bibnamefont {Tiesinga}},
  \bibinfo {author} {\bibfnamefont {C.~J.}\ \bibnamefont {Williams}}, \ and\
  \bibinfo {author} {\bibfnamefont {P.~S.}\ \bibnamefont {Julienne}},\ }\href
  {\doibase 10.1103/PhysRevA.72.042719} {\bibfield  {journal} {\bibinfo
  {journal} {Phys. Rev. A}\ }\textbf {\bibinfo {volume} {72}},\ \bibinfo
  {pages} {042719} (\bibinfo {year} {2005})}\BibitemShut {NoStop}%
\bibitem [{\citenamefont {Fu}\ \emph {et~al.}(2016)\citenamefont {Fu},
  \citenamefont {Li}, \citenamefont {Tey}, \citenamefont {You},\ and\
  \citenamefont {Gao}}]{Gao16a}%
  \BibitemOpen
  \bibfield  {author} {\bibinfo {author} {\bibfnamefont {H.}~\bibnamefont
  {Fu}}, \bibinfo {author} {\bibfnamefont {M.}~\bibnamefont {Li}}, \bibinfo
  {author} {\bibfnamefont {M.~K.}\ \bibnamefont {Tey}}, \bibinfo {author}
  {\bibfnamefont {L.}~\bibnamefont {You}}, \ and\ \bibinfo {author}
  {\bibfnamefont {B.}~\bibnamefont {Gao}},\ }\href {\doibase
  10.1088/1367-2630/18/10/103016} {\bibfield  {journal} {\bibinfo  {journal}
  {New Journal of Physics}\ }\textbf {\bibinfo {volume} {18}},\ \bibinfo
  {pages} {103016} (\bibinfo {year} {2016})}\BibitemShut {NoStop}%
\bibitem [{\citenamefont {Gao}(2004)}]{Gao04a}%
  \BibitemOpen
  \bibfield  {author} {\bibinfo {author} {\bibfnamefont {B.}~\bibnamefont
  {Gao}},\ }\href {http://stacks.iop.org/0953-4075/37/i=11/a=L02} {\bibfield
  {journal} {\bibinfo  {journal} {Journal of Physics B: Atomic, Molecular and
  Optical Physics}\ }\textbf {\bibinfo {volume} {37}},\ \bibinfo {pages} {L227}
  (\bibinfo {year} {2004})}\BibitemShut {NoStop}%
\bibitem [{\citenamefont {Gao}(2005)}]{Gao05b}%
  \BibitemOpen
  \bibfield  {author} {\bibinfo {author} {\bibfnamefont {B.}~\bibnamefont
  {Gao}},\ }\href {\doibase 10.1103/PhysRevLett.95.240403} {\bibfield
  {journal} {\bibinfo  {journal} {Phys. Rev. Lett.}\ }\textbf {\bibinfo
  {volume} {95}},\ \bibinfo {pages} {240403} (\bibinfo {year}
  {2005})}\BibitemShut {NoStop}%
\bibitem [{\citenamefont {Khan}\ and\ \citenamefont {Gao}(2006)}]{KG06}%
  \BibitemOpen
  \bibfield  {author} {\bibinfo {author} {\bibfnamefont {I.}~\bibnamefont
  {Khan}}\ and\ \bibinfo {author} {\bibfnamefont {B.}~\bibnamefont {Gao}},\
  }\href {\doibase 10.1103/PhysRevA.73.063619} {\bibfield  {journal} {\bibinfo
  {journal} {Phys. Rev. A}\ }\textbf {\bibinfo {volume} {73}},\ \bibinfo
  {pages} {063619} (\bibinfo {year} {2006})}\BibitemShut {NoStop}%
\bibitem [{\citenamefont {Li}\ \emph {et~al.}(2016)\citenamefont {Li},
  \citenamefont {Liu}, \citenamefont {Xu}, \citenamefont {de~Melo},\ and\
  \citenamefont {Luo}}]{JL16}%
  \BibitemOpen
  \bibfield  {author} {\bibinfo {author} {\bibfnamefont {J.}~\bibnamefont
  {Li}}, \bibinfo {author} {\bibfnamefont {J.}~\bibnamefont {Liu}}, \bibinfo
  {author} {\bibfnamefont {W.}~\bibnamefont {Xu}}, \bibinfo {author}
  {\bibfnamefont {L.}~\bibnamefont {de~Melo}}, \ and\ \bibinfo {author}
  {\bibfnamefont {L.}~\bibnamefont {Luo}},\ }\href {\doibase
  10.1103/PhysRevA.93.041401} {\bibfield  {journal} {\bibinfo  {journal} {Phys.
  Rev. A}\ }\textbf {\bibinfo {volume} {93}},\ \bibinfo {pages} {041401}
  (\bibinfo {year} {2016})}\BibitemShut {NoStop}%
\bibitem [{\citenamefont {Li}\ \emph {et~al.}(2017)\citenamefont {Li}, ,
  \citenamefont {deMelo},\ and\ \citenamefont {Luo}}]{JL17}%
  \BibitemOpen
  \bibfield  {author} {\bibinfo {author} {\bibfnamefont {J.}~\bibnamefont
  {Li}}, , \bibinfo {author} {\bibfnamefont {L.}~\bibnamefont {deMelo}}, \ and\
  \bibinfo {author} {\bibfnamefont {L.}~\bibnamefont {Luo}},\ }\href {\doibase
  10.3791/55409} {\bibfield  {journal} {\bibinfo  {journal} {J. Vis. Exp.}\
  }\textbf {\bibinfo {volume} {(121)}},\ \bibinfo {pages} {e55409} (\bibinfo
  {year} {2017})}\BibitemShut {NoStop}%
\bibitem [{\citenamefont {Luo}(2008)}]{luothesis}%
  \BibitemOpen
  \bibfield  {author} {\bibinfo {author} {\bibfnamefont {L.}~\bibnamefont
  {Luo}},\ }\emph {\bibinfo {title} {Entropy and Superfluid Critical Parameters
  of a Strongly Interacting Fermi gas}},\ \href@noop {} {Ph.D. thesis},\
  \bibinfo  {school} {Duke University} (\bibinfo {year} {2008})\BibitemShut
  {NoStop}%
\bibitem [{\citenamefont {Gao}(2010)}]{Gao10b}%
  \BibitemOpen
  \bibfield  {author} {\bibinfo {author} {\bibfnamefont {B.}~\bibnamefont
  {Gao}},\ }\href {\doibase 10.1103/PhysRevLett.105.263203} {\bibfield
  {journal} {\bibinfo  {journal} {Phys. Rev. Lett.}\ }\textbf {\bibinfo
  {volume} {105}},\ \bibinfo {pages} {263203} (\bibinfo {year}
  {2010})}\BibitemShut {NoStop}%
\bibitem [{\citenamefont {Gao}(2011)}]{Gao11a}%
  \BibitemOpen
  \bibfield  {author} {\bibinfo {author} {\bibfnamefont {B.}~\bibnamefont
  {Gao}},\ }\href {\doibase 10.1103/PhysRevA.83.062712} {\bibfield  {journal}
  {\bibinfo  {journal} {Phys. Rev. A}\ }\textbf {\bibinfo {volume} {83}},\
  \bibinfo {pages} {062712} (\bibinfo {year} {2011})}\BibitemShut {NoStop}%
\bibitem [{\citenamefont {Gao}(2001)}]{Gao01}%
  \BibitemOpen
  \bibfield  {author} {\bibinfo {author} {\bibfnamefont {B.}~\bibnamefont
  {Gao}},\ }\href {\doibase 10.1103/PhysRevA.64.010701} {\bibfield  {journal}
  {\bibinfo  {journal} {Phys. Rev. A}\ }\textbf {\bibinfo {volume} {64}},\
  \bibinfo {pages} {010701} (\bibinfo {year} {2001})}\BibitemShut {NoStop}%
\bibitem [{\citenamefont {Gao}(2009)}]{Gao09a}%
  \BibitemOpen
  \bibfield  {author} {\bibinfo {author} {\bibfnamefont {B.}~\bibnamefont
  {Gao}},\ }\href {\doibase 10.1103/PhysRevA.80.012702} {\bibfield  {journal}
  {\bibinfo  {journal} {Phys. Rev. A}\ }\textbf {\bibinfo {volume} {80}},\
  \bibinfo {pages} {012702} (\bibinfo {year} {2009})}\BibitemShut {NoStop}%
\bibitem [{\citenamefont {Yan}\ \emph {et~al.}(1996)\citenamefont {Yan},
  \citenamefont {Babb}, \citenamefont {Dalgarno},\ and\ \citenamefont
  {Drake}}]{Yan96}%
  \BibitemOpen
  \bibfield  {author} {\bibinfo {author} {\bibfnamefont {Z.-C.}\ \bibnamefont
  {Yan}}, \bibinfo {author} {\bibfnamefont {J.~F.}\ \bibnamefont {Babb}},
  \bibinfo {author} {\bibfnamefont {A.}~\bibnamefont {Dalgarno}}, \ and\
  \bibinfo {author} {\bibfnamefont {G.~W.~F.}\ \bibnamefont {Drake}},\ }\href
  {\doibase 10.1103/PhysRevA.54.2824} {\bibfield  {journal} {\bibinfo
  {journal} {Phys. Rev. A}\ }\textbf {\bibinfo {volume} {54}},\ \bibinfo
  {pages} {2824} (\bibinfo {year} {1996})}\BibitemShut {NoStop}%
\bibitem [{\citenamefont {Regal}\ \emph {et~al.}(2003)\citenamefont {Regal},
  \citenamefont {Ticknor}, \citenamefont {Bohn},\ and\ \citenamefont
  {Jin}}]{Regal03}%
  \BibitemOpen
  \bibfield  {author} {\bibinfo {author} {\bibfnamefont {C.~A.}\ \bibnamefont
  {Regal}}, \bibinfo {author} {\bibfnamefont {C.}~\bibnamefont {Ticknor}},
  \bibinfo {author} {\bibfnamefont {J.~L.}\ \bibnamefont {Bohn}}, \ and\
  \bibinfo {author} {\bibfnamefont {D.~S.}\ \bibnamefont {Jin}},\ }\href
  {\doibase 10.1103/PhysRevLett.90.053201} {\bibfield  {journal} {\bibinfo
  {journal} {Phys. Rev. Lett.}\ }\textbf {\bibinfo {volume} {90}},\ \bibinfo
  {pages} {053201} (\bibinfo {year} {2003})}\BibitemShut {NoStop}%
\bibitem [{\citenamefont {Ticknor}\ \emph {et~al.}(2004)\citenamefont
  {Ticknor}, \citenamefont {Regal}, \citenamefont {Jin},\ and\ \citenamefont
  {Bohn}}]{Ticknor04}%
  \BibitemOpen
  \bibfield  {author} {\bibinfo {author} {\bibfnamefont {C.}~\bibnamefont
  {Ticknor}}, \bibinfo {author} {\bibfnamefont {C.~A.}\ \bibnamefont {Regal}},
  \bibinfo {author} {\bibfnamefont {D.~S.}\ \bibnamefont {Jin}}, \ and\
  \bibinfo {author} {\bibfnamefont {J.~L.}\ \bibnamefont {Bohn}},\ }\href
  {\doibase 10.1103/PhysRevA.69.042712} {\bibfield  {journal} {\bibinfo
  {journal} {Phys. Rev. A}\ }\textbf {\bibinfo {volume} {69}},\ \bibinfo
  {pages} {042712} (\bibinfo {year} {2004})}\BibitemShut {NoStop}%
\bibitem [{\citenamefont {Cui}\ \emph {et~al.}(2017)\citenamefont {Cui},
  \citenamefont {Shen}, \citenamefont {Deng}, \citenamefont {Dong},
  \citenamefont {Chen}, \citenamefont {L\"u}, \citenamefont {Gao},
  \citenamefont {Tey},\ and\ \citenamefont {You}}]{Gao17b}%
  \BibitemOpen
  \bibfield  {author} {\bibinfo {author} {\bibfnamefont {Y.}~\bibnamefont
  {Cui}}, \bibinfo {author} {\bibfnamefont {C.}~\bibnamefont {Shen}}, \bibinfo
  {author} {\bibfnamefont {M.}~\bibnamefont {Deng}}, \bibinfo {author}
  {\bibfnamefont {S.}~\bibnamefont {Dong}}, \bibinfo {author} {\bibfnamefont
  {C.}~\bibnamefont {Chen}}, \bibinfo {author} {\bibfnamefont {R.}~\bibnamefont
  {L\"u}}, \bibinfo {author} {\bibfnamefont {B.}~\bibnamefont {Gao}}, \bibinfo
  {author} {\bibfnamefont {M.~K.}\ \bibnamefont {Tey}}, \ and\ \bibinfo
  {author} {\bibfnamefont {L.}~\bibnamefont {You}},\ }\href {\doibase
  10.1103/PhysRevLett.119.203402} {\bibfield  {journal} {\bibinfo  {journal}
  {Phys. Rev. Lett.}\ }\textbf {\bibinfo {volume} {119}},\ \bibinfo {pages}
  {203402} (\bibinfo {year} {2017})}\BibitemShut {NoStop}%
\bibitem [{\citenamefont {Yao}\ \emph {et~al.}()\citenamefont {Yao},
  \citenamefont {Qi}, \citenamefont {Liu}, \citenamefont {Wang}, \citenamefont
  {Wang}, \citenamefont {Wu}, \citenamefont {Chen}, \citenamefont {Zhang},
  \citenamefont {Zhai}, \citenamefont {Chen},\ and\ \citenamefont
  {Pan}}]{Yao17}%
  \BibitemOpen
  \bibfield  {author} {\bibinfo {author} {\bibfnamefont {X.-C.}\ \bibnamefont
  {Yao}}, \bibinfo {author} {\bibfnamefont {R.}~\bibnamefont {Qi}}, \bibinfo
  {author} {\bibfnamefont {X.-P.}\ \bibnamefont {Liu}}, \bibinfo {author}
  {\bibfnamefont {X.-Q.}\ \bibnamefont {Wang}}, \bibinfo {author}
  {\bibfnamefont {Y.-X.}\ \bibnamefont {Wang}}, \bibinfo {author}
  {\bibfnamefont {Y.-P.}\ \bibnamefont {Wu}}, \bibinfo {author} {\bibfnamefont
  {H.-Z.}\ \bibnamefont {Chen}}, \bibinfo {author} {\bibfnamefont
  {P.}~\bibnamefont {Zhang}}, \bibinfo {author} {\bibfnamefont
  {H.}~\bibnamefont {Zhai}}, \bibinfo {author} {\bibfnamefont {Y.-A.}\
  \bibnamefont {Chen}}, \ and\ \bibinfo {author} {\bibfnamefont {J.-W.}\
  \bibnamefont {Pan}},\ }\href@noop {} {}\Eprint
  {http://arxiv.org/abs/arXiv:1711.06622} {arXiv:1711.06622} \BibitemShut
  {NoStop}%
\end{thebibliography}


%

\end{document}